\documentclass[%
reprint,
superscriptaddress,
showpacs,
 amsmath,amssymb,
 aps,
 prl,
]{revtex4-1}

\usepackage{graphicx}
\usepackage{dcolumn}
\usepackage{bm}
\usepackage{amsmath}
\usepackage{autobreak}
\usepackage{hyperref}
\usepackage[mathlines]{lineno}
\usepackage{natbib}
\usepackage{multirow}
\usepackage{booktabs}
\usepackage{xspace}
\usepackage{float}
\usepackage{diagbox}
\usepackage{subfigure}

\begin{document}

\preprint{APS/123-QED}

\title{Test of electric charge conservation in dark matter direct detection experiment}

\author{Z.~H.~Zhang}
\email{Corresponding author: zhenhua@bnu.edu.cn}
\affiliation{Beijing Normal University at Zhuhai, Zhuhai 519087, China}
\author{X.~P.~Geng}
\email{Corresponding author: gxp18@tsinghua.org.cn}
\affiliation{Institute of Applied Physics and Computational Mathematics, Beijing 100094, China}
\author{J.~W.~Hu}
\affiliation{Beijing Honest Technology Co., Ltd., Beijing, 100083, China}
\author{Z.~H.~Zhang}
\affiliation{China Nuclear Power Technology Research Institute Co., Ltd., Shenzhen, 518000, China}

\date{\today}

\begin{abstract}
Electric charge conservation (ECC) is typically taken as an axiom in the standard model. Searching for small violations with high-performance experiments could lead us to new physics. In this work, we tested ``invisible" electric charge nonconservation (ECNC) events with 1.16 ton$\cdot$year electron recoil data from the XENON-nT experiment. There was no  statistically significant signal, and the ECNC limit was updated to $\tau(e^-\rightarrow{\nu_e+\bar{\nu}_e+\nu_e}) > 4.34 \times10^{27}$ yr. This work increases the limit by two orders of magnitude and shows that dark matter direct detection experiments have great potential for further ECC testing.
\end{abstract}

\maketitle

\emph{Introduction.}
In 1957, Wu et al. reported that parity is not conserved in weak interactions in $\beta$ decay experiments~\cite{WuCS}. In 1964, Christenson et al. discovered the violation of charge-parity invariance~\cite{CPV1964}. These findings greatly shocked the community. Since then, new experiments have been conducted to examine fundamental laws, such as electric charge conservation (ECC).

ECC arises from the global $U(1)$ symmetry gauge invariance of the electric field~\cite{Noether1918}. The experimental evidence for ECC is that a photon has an extremely small upper limit on its rest mass~\cite{Weinberg_1964}. However, electric charge nonconservation (ECNC) is allowed in some theoretical frameworks beyond the standard model~\cite{Witten,CHU,IGNATIEV1979,Sergei2000}. If ECNC is confirmed, then new physics will emerge. However, to date, no definite signal of ECNC has been found. In previous tests of ECC, the limits on the mean lifetime $\tau$ have gradually become increasingly stringent. The most recent result of $\tau(e^-\rightarrow{\nu_e+\bar{\nu}_e+\nu_e}) > 2.83 \times10^{25}$ yr was from the MAJORANA DEMONSTRATOR with a 37.5 kg yr exposure~\cite {MJD2024}.

The MAJORANA experiment pursues neutrinoless double-beta decay via a high-purity germanium (HPGe) array with an ultralow background~\cite{MAJORANAExperiment}. Dark matter direct detection experiments such as XENON~\cite{XENONnT_EPJC2024,MJD_0vbb_2023,MJD_EDM_2024} and PandaX~\cite{PandaX_PRL2023} also have an ultralow background, can carry out the search for the other ultrarare physical events while searching for dark matter~\cite{MJD_EDM_2024}.

The event rate is approximately 0.06 Count/keV/kg/day, which is 21900 Count/keV/ton/yr at 5 keV, approximately 8000 Count/keV/ton/yr at 11 keV in the MAJORANA DEMONSTRATOR experiment~\cite{MJD2024}, and approximately 12 Count/keV/ton/yr at 5 keV in the XENON-nT experiment~\cite{XENONnT_ER2022}. HPGe detector usually has an excellent energy resolution of approximately 80 eV at 5 keV. Although the energy resolution of XENON-nT (approximately 700 eV at 5 keV) can not match for that of the MAJORANA DEMONSTRATOR, XENON-nT has a lower event rate, which allows it to obtain better results. In this work, we tested the ECC with 1.16 ton$\cdot$year exposure data from the XENON-nT experiment and updated the most stringent limit.

\emph{Expected signal}

ECNC occurs when an electron decays into neutral particles. There are two main types of experiments on ECNC in electron decay:

I. ``visible", $e^-\rightarrow{\nu_e+\gamma}$. This process splits the energy of the electrons equally, giving them 256 keV each. The photon can be picked up by the detector, so this process is ``visible". BOREXINO is a typical example of this type of experiment~\cite{Borexino2015,Borexino2017}.

II. ``invisible", $e^-\rightarrow{\nu_e+\bar{\nu}_e+\nu_e}$, the most favorable mode~\cite{Workman_2022}. This process produces neutrinos instead of photons, so it is called ``invisible" decay of the electron. When one of the inner electrons of an atom is lost by invisible decay, the outer electron fills the electron vacancy, and X-rays are be emitted.The atomic levels of xenon are listed in Table~\ref{tab::TABLE I}. For example, when an electron in the $K$ shell undergoes invisible decay in a Xe atom, X-rays of 34.6 keV are detected. In this work, we choose this most favorable mode to test ECC and search for ECNC.

\begin{table} [htbp]
	\centering
	\caption{\textbf{Electron binding energies of xenon atom.}}
	\label{tab::TABLE I}
	\begin{tabular}{lcc} 
		\\
		\hline
		Shell & Energy (keV) & Electron Number \\
		\hline
          $K~1s$ & $34.6$ & $2$ \\
          $L_1~2s$ & $5.45$ & $2$ \\
          $L_2~2p_{1/2}$ & $5.11$ & $2$ \\
          $L_3~2p_{3/2}$ & $4.79$ & $4$ \\
          $M_1~3s$ & $1.15$ & $2$ \\
          $M_2~3p_{1/2}$ & $1.00$ & $2$ \\
          $M_3~3p_{3/2}$ & $0.94$ & $4$ \\
          $M_4~3d_{3/2}$ & $0.69$ & $4$ \\
          $M_5~3d_{5/2}$ & $0.68$ & $6$ \\
          others & $<0.3$ & $26$ \\
		\hline
	\end{tabular}
\end{table}

Considering the energy resolution of the detector and the total efficiency $\eta_\mathrm{eff} (E)$ of the experiment, the ECNC signal $S$ in the experiment can be described by Eq.~\ref{eq::eq2}. For the sake of expression, we define the signal that does not take $\eta_\mathrm{eff}$ into account as a physical signal $P$, and the signal that takes $\eta_\mathrm{eff}$ into account as a detected signal $S$.

\begin{equation}
	S = P\cdot \eta_\mathrm{eff} = 
    \sum_i \frac{I_{i}}{\sqrt{2\pi}\sigma_{i}}
    \mathrm{exp}{\left[-\frac{(E-E_{i})^2}{2\sigma_{i}^2}\right]}
    \cdot \eta_\mathrm{eff},
	\label{eq::eq2} 
\end{equation}
where, $I_{i}$ and $\sigma_{i}$ are the intensity and the energy resolution (standard deviation) of the $i$-th peak, respectively. 

\begin{figure} [htb]
	\centering
	\includegraphics[width=0.99\linewidth]{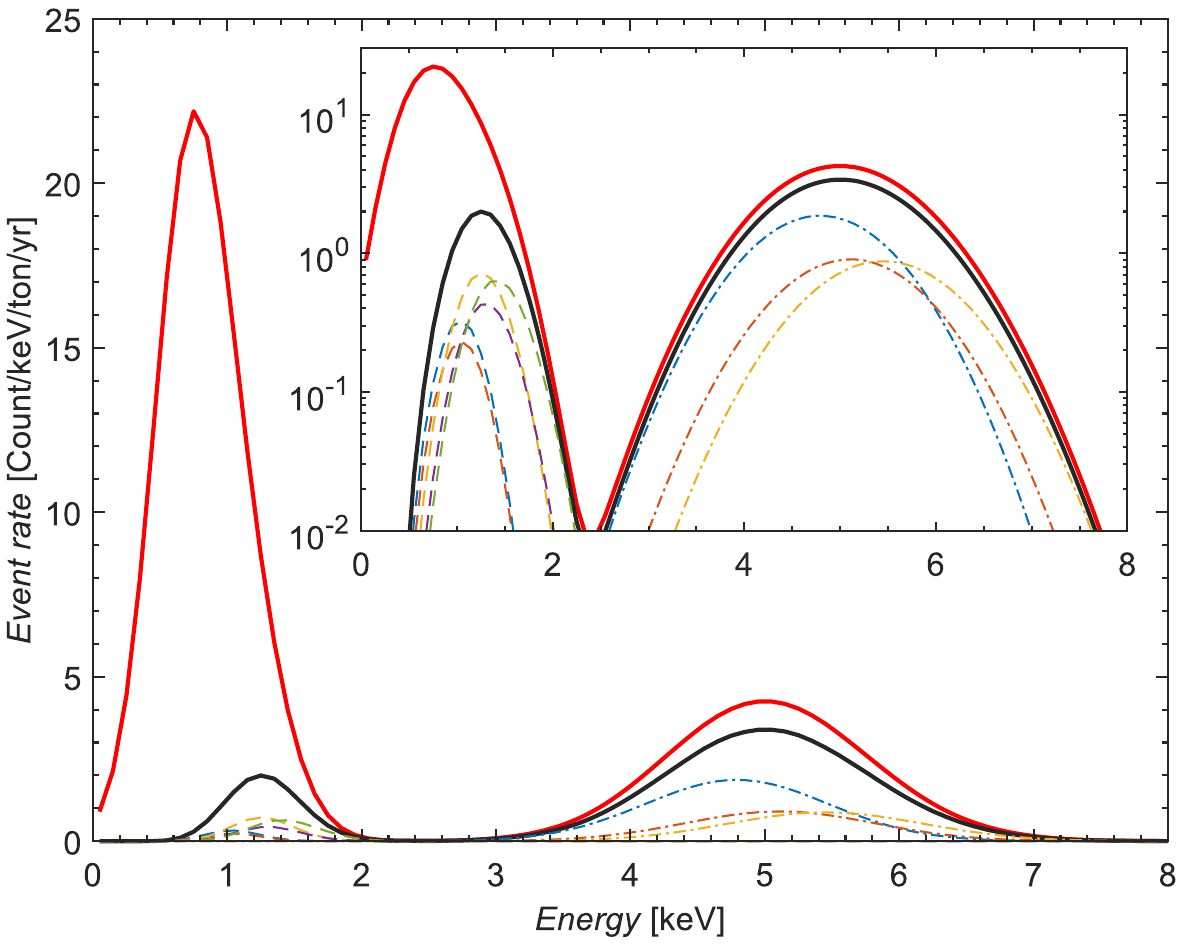} 
	\caption{Expected signal of the ``invisible'' ECNC in the XENON-nT detector. The physical and detected signals are plotted as red and black solid lines, respectively. The large peaks of the detected signal contain several smaller peaks (dashed and dot-dashed lines).The total intensity of ECNC is 26 Count/ton/yr; for more details, see the text.}
	\label{fig::ECNCPhysicalSpec} 
\end{figure}

In the example shown in Fig.~\ref{fig::ECNCPhysicalSpec}, the total intensity of ECNC is $I \equiv \sum I_i = 26$ Count/ton/yr. We distribute the total intensity to the peaks in proportion to the number of electrons as listed in Table~\ref{tab::TABLE I},$I_1(0.68~\mathrm{keV}):I_2(0.69~\mathrm{keV}):I_3(0.94~\mathrm{keV}):I_4(1.00~\mathrm{keV}):I_5(1.15~\mathrm{keV}):I_6(4.79~\mathrm{keV}):I_7(5.11~\mathrm{keV}):I_8(5.45~\mathrm{keV}) = 6:4:4:2:2:4:2:2$.
Owing to $\eta_\mathrm{eff}$, the physical signal represented by the thick red line moves down to the expected signal in the xenon detector represented by the black line. The large peaks of the expected signal contain several smaller peaks (thin dotted and dash-dotted lines). The energy resolution $\sigma \mathrm{(keV)} = 0.310 \sqrt{E \mathrm{(keV)}} +0.0037$ is referenced from XENON-1T~\cite{XENON1T_ER2020, XENONnT_ER2022}, and $\eta_\mathrm{eff}$ is derived from public data on XENON's website~\cite{XENON_Website}.

\emph{Results and discussion}

The expected energy spectrum consists of the detected signal ($S$) and the background ($B$). The expected spectrum $\left(S+B\right)$ and the measured spectrum $n$ were compared via minimum-$\chi^2$ analysis:

\begin{equation}
\chi^2 = \sum_j \frac{\left[n_j-\left(S_j+B_j\right)\right]^2}{\Delta^2_j},
\label{eq::eq4}
\end{equation}
where $n_j$ and $\Delta _j$ denote the measured data and standard deviation with statistical and systematical components at the $j$-th energy bin extracted from Fig. 5 in Ref.~\cite{XENONnT_ER2022}, respectively; $S_j$ denotes the ECNC event rate at the $j$-th energy bin; and $B_j$ denotes the background contribution at the $j$-th energy bin in Ref.~\cite{XENONnT_ER2022, XENON_Website}. We found that the no-signal hypothesis yielded a $\chi^2_0 = 15.91$. The one-sided 90\% CL limit was estimated with $\chi^2 = \chi^2_0 + \Delta \chi^2 = 18.62$, based on Feldman-Cousins unified approach~\cite{FCChiSquare}.

By Eq.~\ref{eq::ItoTau}, the event intensity $I$ can be converted into the mean lifetime $\tau$.

\begin{equation}
\tau = \frac{n_e\cdot N}{I},
\label{eq::ItoTau}
\end{equation}
where $N = 4.58\times10^{27}$ ton$^{-1}$ is the number of Xe atoms  per ton, and $n_e$ is the number of tested electrons in an atom.

\begin{figure} [h]
	\centering
	\includegraphics[width=0.99\linewidth]{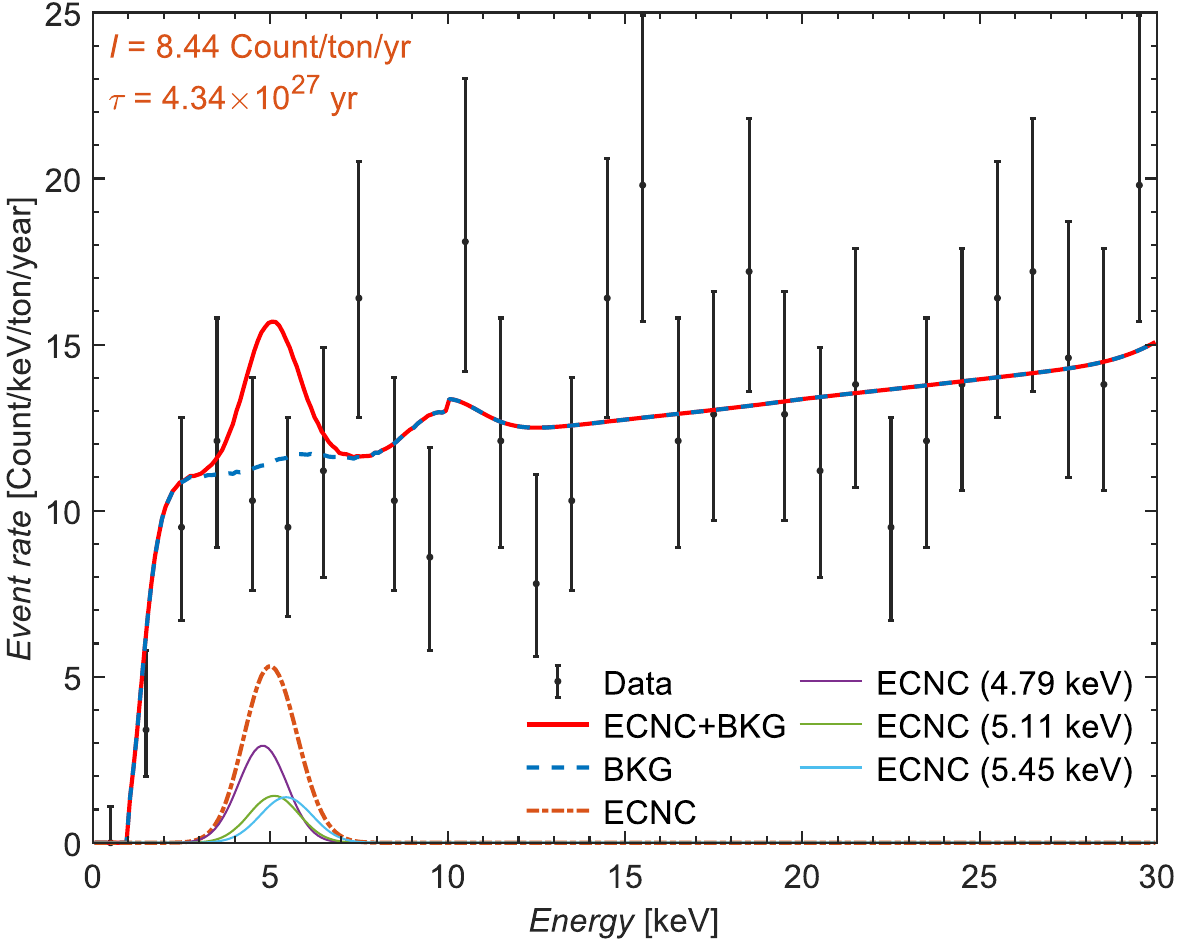} 
	\caption{Matching between the experimental and expected energy spectra. The data points with error bars represent the electron recoil spectrum in the XENON-nT experiment, extracted from Fig. 5 in Ref.~\cite{XENONnT_ER2022}. The BKG (blue dashed line) is the background $B_0$ in Ref.~\cite{XENONnT_ER2022} and Web.~\cite{XENON_Website}. Three peaks attributed to the 8 $L$ shell electrons are plotted with thin solid lines.}
	\label{fig::SpectraMatching} 
\end{figure}

As shown in Fig.1, the $M$ shell with more electrons contributes less than the L shell does, so we chose to test only the 8 $L$ shell electrons (corresponding to $I_6$, $I_7$ and $I_8$) to derive a more conservative and solid result. We did not test the $K$ shell electrons because not enough experimental data are currently available. There was no statistically significant signal in the experimental spectrum~\cite{XENONnT_ER2022}, as shown by the data points with error bars in Fig.~\ref{fig::SpectraMatching}. When $I \geq 8.44$ Count/ton/yr, $\chi^2 \geq 18.62$, as shown by the thick red line in Fig.~\ref{fig::SpectraMatching}. Thus, we can set a 90\% CL limit of ECNC in ``invisible'' electron decay: $\tau(e^-\rightarrow{\nu_e+\bar{\nu}_e+\nu_e}) > 4.34\times10^{27}$ yr.

\begin{figure} [h]
	\centering
	\includegraphics[width=0.99\linewidth]{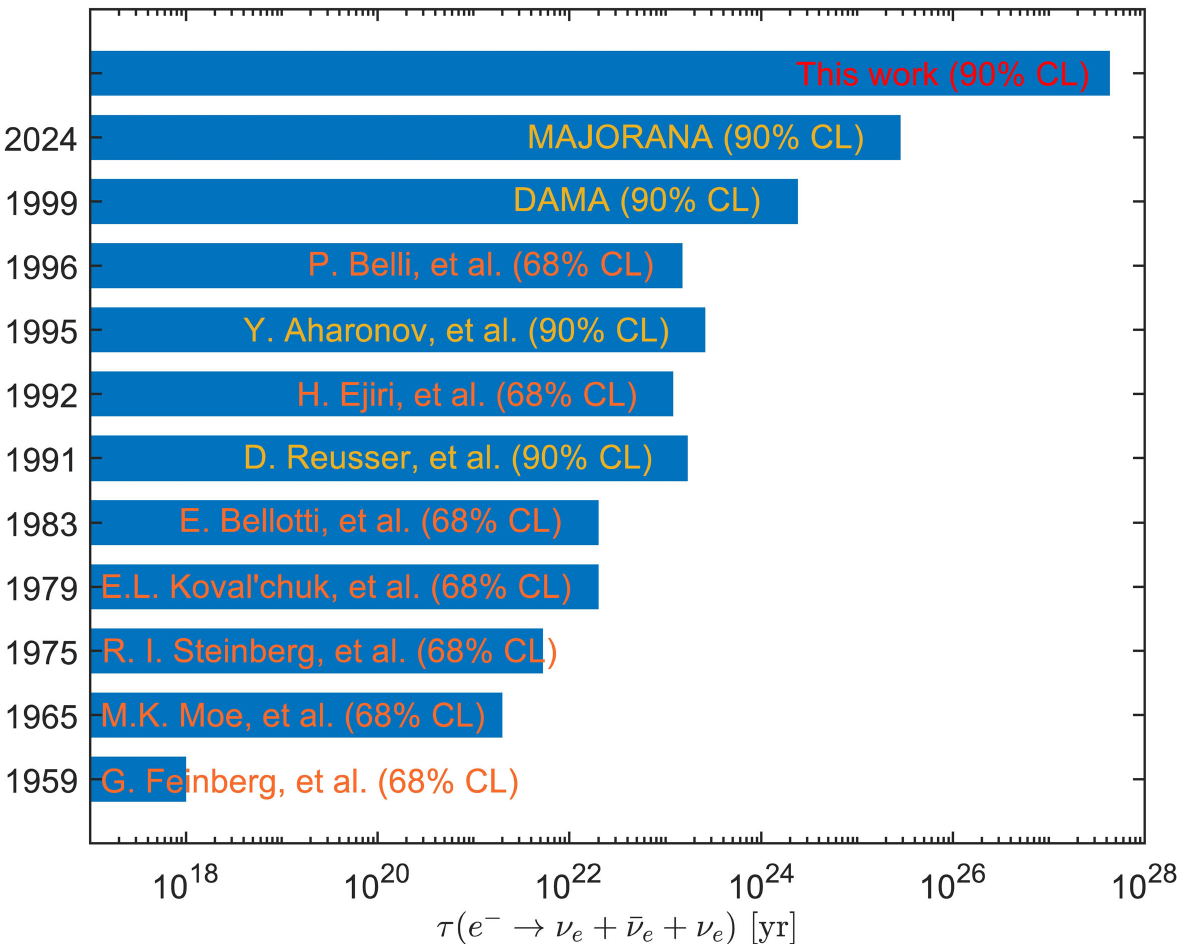} 
	\caption{The lower limits on the mean lifetime of ECNC in electron decay $e^-\rightarrow{\nu_e+\bar{\nu}_e+\nu_e}$. The previous results are from Ref.~\cite{Feinberg_1959, Moe_1965, Steinberg_1975, Kovalchuk_1979, Bellotti_1983, Reusser_1991, Ejiri_1992, Aharonov_1995, Belli_1996, Belli_1999, MJD2024}.}
	\label{fig::ResultComparison} 
\end{figure}

As shown in Fig.~\ref{fig::ResultComparison}, starting in 1959, researchers and collaborations have continually pushed the lower limit. The pace of updating results was relatively slow in the 1990s until the DAMA experiment in 1999 gave significantly stronger limits. After more than 20 years of silence, MAJORANA again significantly refreshed the limits. Finally, this work has rapidly increased the limit by two orders of magnitude with XENON-nT's 1.16 ton$\cdot$year electron recoil data.

This work, together with the DAMA and MAJORANA's explorations, reveals the advantages of dark matter detection experiments to examine ECC and search for ECNC cases. Experiments such as XENON and MAJORANA, which have ultralow background and excellent energy resolution, have great potential for further ECC testing and searching for new physical ECNC with their greater exposure.

\bibliography{ECNC}

\end{document}